\documentclass{article}
\usepackage{amsmath}
\usepackage{amssymb}
\usepackage{graphics}
\usepackage{caption}
\usepackage{geometry}
\usepackage{media9}
\begin{document}
\begin{center}
{\bf\Large Rigorous analytical formula for freeform singlet lens design free of spherical aberration}\\[5mm]

\noindent Rafael G. Gonz\'alez-Acu\~na$^{1,*}$, H\'ector A. Chaparro-Romo$^{2}$\\ and 
Julio C. Guti\'errez-Vega$^{1}$\\[5mm]

\begin{minipage}{0.5\textwidth}
\noindent $^{1}$ Photonics and Mathematical Optics Group,\\ Tecnol\'ogico de Monterrey, Monterrey 64849, M\'exico\\
$^{2}$UNAM, Av. Universidad 3000, Cd. Universitaria, Coyoac\'an, 04510 Ciudad de M\'exico, M\'exico.\\

{\it email} * rafael123.90@hotmail.com\\ 
\end{minipage}

{\bf Abstract}\\[5mm]

\begin{minipage}{0.9\textwidth}
An analytical closed-form formula for the design of freeform lenses free of spherical aberration is presented. Given the equation of the freeform input surface, the formula gives the equation of the second surface in order to correct the spherical aberration. The derivation is based on the formal application of the variational Fermat principle under the standard geometrical optics approximation.
\end{minipage}\end{center}

\section{Introduction} 
Freeform optics involves the design of optical elements with at least one surface which has no translational or rotational symmetry about a propagation axis. In recent years, the topic has gained increasing popularity in the optics community, partly because of the rapid development of new computing technologies and the emergence of potential applications. In general, the design of freeform elements has combined theoretical approximation methods with brute-force optimization techniques leading to a diversity of results and methodologies which have proved to be useful for particular cases \cite{bauer2018starting,yang2017automated}. For instance, G. W. Forbes \cite{forbes2012characterizing, forbes2013fitting, forbes2010robust, forbes2007shape} described freeform surfaces based on a set of characteristic polynomials for non rotationally symmetric systems. Recently  the generation of freeform mirrors have been studied \cite{muslimov2017combining, bauer2015design}, which considerably reduce the optical aberrations. 
The theory of aberration of freeform optics has been developed by several authors \cite{ochse2018aberration, fuerschbach2014theory, zhong2018vectorial, zhong2017initial} applying numerical optimization schemes. 

In this paper, we introduce a closed-form expression for the design of freeform singlets lenses free of spherical aberration, which is a continuation of our work \cite{gonzalez2018general,gonzalez2018generalization}. The formula gives the exact analytical equation of the output surface given the arbitrary freeform expresion of the input surface in order to correct the spherical aberration introduced by the first surface. The derivation is fully analytical based on the formal application of the variational Fermat principle under the standard geometrical optics approximation. In the process of deriving the formula, we apply a design methodology free of numerical optimization strategies. We illustrate the applicability and robustness of the formula by showing some representative design examples using very sophisticated input functions that have no been used before in optical design. As far as we know, this exact formula has not been reported before in the optical design literature. 

\section{Analytical design of freeform singlet free of spherical aberration}

We assume that the singlet lens is a lossless and homogeneous optical element with relative refractive index $n$ and axial thickness $T$, see Fig. 1. Its input surface is known and it is described by the freeform function $z_{a}(x_{a},y_{a})$, where the subindex $a$
refers to the coordinates on the input surface. The shape of the output surface is unknown and it is described by the function $z_{b}(x_{b},y_{b})$ to be determined, where the subindex $b$ refers to the coordinates on the output surface. We will further assume that the normal vector of the input surface at the optical axis points out in direction $z$, i.e. the normal is perpendicular to the tangent plane of the input surface at the origin.

The goal is to determine the output function $z_{b}(x_{b},y_{b})$ given the input function $z_{a}(x_{a},y_{a})$ in order to correct the spherical aberration introduced by the first surface.
To do this, we will consider an object point (O) located at $z=f_{a}$ on the
optical axis and its corresponding image point (I) focused at $z=T+f_{b}$, as shown in Fig. 1(a). The input field impinging on the freeform singlet is a monochromatic spherical
wave emerging from the source (O). Since the size of the freeform singlet is
much larger than the wavelength of the light, a ray optics representation may
be applied to solve this problem. In this approach, the input field is
characterized by an uniform bundle of radial rays emerging from (O).

\begin{center}
\begin{figure}[h]
\begin{center}
\includegraphics{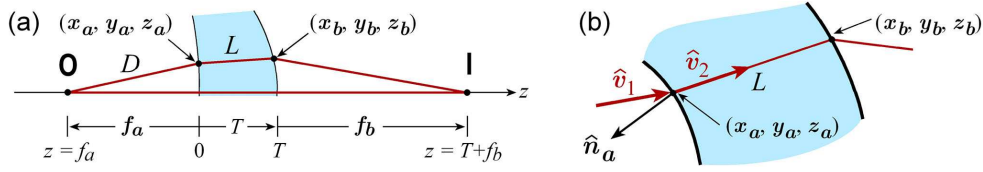}
\caption{(a) Geometry of the problem and notation used for the distances. The origin of the coordinate system is located at the center of the input surface $z_a\left(  0,0\right)=0.$ (b) Zoom showing the notation for the unit vectors.}
\end{center}
\end{figure}
\end{center}

We begin by recalling the Snell law in vector form for an incident ray coming from medium $n_1$ and transmitted to medium $n_2$ \cite{glassner1989introduction}
\begin{equation}
{n\,\hat{\mathbf{v}}_{2}=[\hat{\mathbf{n}}_{a}\times(-\hat{\mathbf{n}}_{a}\times\hat{\mathbf{v}}_{1})]-\,\hat{\mathbf{n}}_{a}\sqrt{n^{2}-(\hat{\mathbf{n}}_{a}\times\hat{\mathbf{v}}_{1})\cdot(\hat{\mathbf{n}}_{a}\times\hat{\mathbf{v}}_{1})}},\label{eq:snell0}%
\end{equation}
where $n=n_2/n_1$ is the relative refractive index, $\hat{\mathbf{n}}_{a}$ is the normal unit vector of the input surface pointing towards the incident medium, and $\hat{\mathbf{v}}_{1}$, $\hat{\mathbf{v}}_{2}$ are the propagation unit vectors of the incident and refracted rays, respectively, see Fig. 1(b). If the ray strikes the input surface at the point $(x_{a},y_{a},z_{a})$ and emerges from the lens at $(x_{b},y_{b},z_{b})$, then the unit vectors can be expressed as
\begin{equation}
{\hat{\mathbf{n}}_{a}=\frac{[z_{a_{x}},z_{a_{y}},-1]}{S}}\,,\quad
{\hat{\mathbf{v}}_{1}=\frac{[x_{a},y_{a},z_{a}-f_{a}]}{D}}\,,\quad
{\hat{\mathbf{v}}_{2}=\frac{[x_{b}-x_{a},y_{b}-y_{a},z_{b}-z_{a}]}{L}%
},\label{eq:vec1}%
\end{equation}
where%
\begin{equation}
S\equiv\sqrt{{z_{a_{x}}^{2}\hspace{-0.5mm}+\hspace{-0.5mm}z_{a_{y}}^{2}%
\hspace{-0.5mm}+\hspace{-0.5mm}1}},\quad D\equiv\sqrt{{x_{a}^{2}%
\hspace{-0.5mm}+\hspace{-0.5mm}y_{a}^{2}\hspace{-0.5mm}+\hspace{-0.5mm}\left(
z_{a}\hspace{-0.6mm}-\hspace{-0.6mm}f_{a}\right)  ^{2}}},\quad L\equiv
\sqrt{{\left(  x_{b}\hspace{-0.6mm}-\hspace{-0.6mm}x_{a}\right)  ^{2}%
\hspace{-0.5mm}+\hspace{-0.5mm}\left(  y_{b}\hspace{-0.6mm}-\hspace
{-0.6mm}y_{a}\right)  ^{2}\hspace{-0.5mm}+\hspace{-0.5mm}\left(  z_{b}%
\hspace{-0.6mm}-\hspace{-0.6mm}z_{a}\right)  ^{2}}},
\end{equation}
and $z_{a_{x}}\equiv\partial_{x}z_{a}$ and $z_{a_{y}}\equiv\partial_{y}z_{a}$
are the partial derivatives of $z_{a}\left(  x_{a},y_{a}\right)  $ with
respect to $x_{a}$ and $y_{a}$. Note that $S$ and $D$ depend on the input
surface exclusively, whereas $L$ depends on both input and output surfaces.

Replacing the unit vectors Eqs. (\ref{eq:vec1}) into the vector equation (\ref{eq:snell0}) and separating the Cartesian components we
get the following expressions for the direction cosines $\mathcal{X},\mathcal{Y},\mathcal{Z}$ of the vector ${\hat{\mathbf{v}}_{2}}$
\begin{subequations}
\label{eq:vecsnell}%
\begin{align}
\mathcal{X} &  \equiv{\frac{x_{b}-x_{a}}{L}=\frac{x_{a}\left(  z_{a_{y}}^{2}+1\right)  -z_{a_{x}}\left(  y_{a}z_{a_{y}}+f_{a}-z_{a}\right)  }{nDS^{2}}-z_{a_{x}}}\frac{{\Phi}}{S},\\
\mathcal{Y} &  \equiv{\frac{y_{b}-y_{a}}{L}=\frac{y_{a}\left(  z_{a_{x}}^{2}+1\right)  -z_{a_{y}}\left(  x_{a}z_{a_{x}}+f_{a}-z_{a}\right)  }{nDS^{2}}-z_{a_{y}}}\frac{{\Phi}}{S},\\
\mathcal{Z} &  \equiv{\frac{z_{b}-z_{a}}{L}=\frac{\left(z_{a}-f_{a}\right)
\left(  z_{a_{x}}^{2}+z_{a_{y}}^{2}\right)  +x_{a}z_{a_{x}}+y_{a}z_{a_{y}}%
}{nDS^{2}}+}\frac{{\Phi}}{S},
\end{align}
\end{subequations}
where
\begin{equation}
{\Phi\equiv\left[  {1-\frac{\left(  y_{a}z_{a_{x}}-x_{a}z_{a_{y}}\right)
^{2}+\left[  z_{a_{x}}\left(  z_{a}-f_{a}\right)  +x_{a}\right]  ^{2}+\left[
z_{a_{y}}\left(  z_{a}-f_{a}\right)  +y_{a}\right]  ^{2}}{n^{2}D^{2}S^{2}}}\right]  }^{1/2},
\end{equation}
and, evidently,  $\mathcal{X}^2 + \mathcal{Y}^2 + \mathcal{Z}^2 = 1$. 
Relations (\ref{eq:vecsnell}) come from the application of the Snell law for an arbitrary ray striking the singlet lens at point $(x_a,y_a,z_a)$. Note that the expressions in the right sides of Eqs. (\ref{eq:vecsnell}) are fully expressed in terms of the coordinates of the input surface, that is, $\mathcal{X}=\mathcal{X}(x_a,y_a,z_a)$ and so on. 

We will now to derive an additional relation in order to fulfill the free
spherical aberration condition. Let us consider two rays emerging simultaneously from (O), the first one
strikes the singlet at $(x_{a},y_{a},z_{a})$ and the second one travels along
the optical axis, see Fig. 1(a). Both rays pass through the singlet and meet
again at the image point (I) located at $(0,0,T+f_{b})$. The Fermat principle requires that both optical lengths between points (O) and (I) be the same, thus equating the optical paths we get%
\begin{equation}
-f_{a}+nT+f_{b}=-\,{\textrm{sgn}(f_{a})D\ +\ nL+\textrm{sgn}(f_{b})\sqrt{x_{b}%
^{2}+y_{b}^{2}+(z_{b}-T-f_{b})^{2}},}\label{fer}%
\end{equation}
where $\textrm{sgn}(\bullet)$ is the sign function. For negative (positive) values of $f_{a}$ the object is real
(virtual), and positive (negative) values of $f_{b}$ the image is real (virtual).

Equations (\ref{eq:vecsnell}) and (\ref{fer}) form a system of algebraic
equations for the unknowns $x_{b}$, $y_{b}$ and $z_{b}$, whose exact
solution is given by
\begin{equation}
\label{sol}
x_{b}={x_{a}+\frac{\mathcal{X}(z_{b}-z_{a})}{\mathcal{Z}}},\qquad 
y_{b}={y_{a}+\frac{\mathcal{Y}(z_{b}-z_{a})}{\mathcal{Z}}},\qquad 
z_{b}={{\frac{g-\sqrt{g^2+h\left(  n^2-1\right) }}{{n^2-1}}}},
\end{equation}
where%
\begin{subequations}
\begin{align}
g & \equiv \left(  z_{a}-f_{b}-T\right)\mathcal{Z}^{2}+q\mathcal{Z}+z_{a}\left(n^{2}-1\right),\\
h & \equiv\left[  x_{a}^{2}+y_{a}^{2}-z_{a}^{2}+\left(  T+f_{b}\right)^{2}-\left(  p-nT\right)  ^{2}\right]  {\mathcal{Z}^{2}-2z_{a}q\mathcal{Z-}z_{a}^{2}\left(  n^{2}-1\right)},\\
p & \equiv{-\mathrm{sgn}(f_{a})D+f_{a}-f_{b},}\\
q & \equiv x_{a}\mathcal{X}+y_{a}\mathcal{Y}-np+n^{2}T.
\end{align}
\end{subequations}

Equations (\ref{sol}) are the most important result of this paper. They describe analytically the shape $z_b(x_b,y_b)$ of the output surface of the singlet lens in terms of the function $z_a(x_a,y_a)$ of its freeform input surface and the design parameters $(f_a,f_b,n,T)$. These expressions may look cumbersome, but it is quite remarkable that could be expressed in closed-form for an arbitrary freeform input surface. As far as we know, these relations have not been derived before. We recall that a necessary condition for the validity of Eqs. (\ref{sol}) is that the surface normal should be perpendicular to the tangent plane to the input surface at the origin. 

From a more mathematical point of view, since the freeform lens is an homogeneous optical element, the input and output surfaces are simple connected sets on $\mathbb{R}^3$ that can be defined as
\begin{equation}
\Psi_a=\{(x_a,y_a,z_a)\in \mathbb{R}^3| z_a<z_b  \},\hspace{1cm}
\Psi_b=\{(x_b,y_a,z_b)\in \mathbb{R}^3| z_b>z_a  \},
\end{equation}
where $\Psi_a$ and $\Psi_b$ are homeomorphic, which means that both surfaces are topologically equivalent. Thus, there exists a continuous and bijective function $f$ such that $f:\Psi_a \rightarrowtail \hspace{-1.5ex} \rightarrow \Psi_b$, and whose inverse $f^{-1}$ is also continuous. There are many functions $f$ that map both sets, but there is only one that is physically valid and corresponds to that one which satisfies the variational Fermat principle of minimum optical length. In our case, it is clear that $f$ is given by Eqs. (\ref{sol}).  The uniqueness of $f$ has as consequence that the Snell law is automatically fulfilled at the second interface $z_b$ as well. Now, since $f$ is continuous it means that $f$ maps open balls from $\Psi_a$ to $\Psi_b$, then the ray neighborhoods are preserved. Therefore, the validity of Eqs. (\ref{sol}) also requires that the rays do not intersect each other inside the lens because, in this  case, $\Psi_b$ overlaps itself leaving from being homeomorphic with respect to $\Psi_a$, and the vicinity of the neighborhoods are not preserved.\\

\begin{center}
\begin{figure}[h]
\begin{center}
\resizebox{150mm}{90mm}{\includegraphics{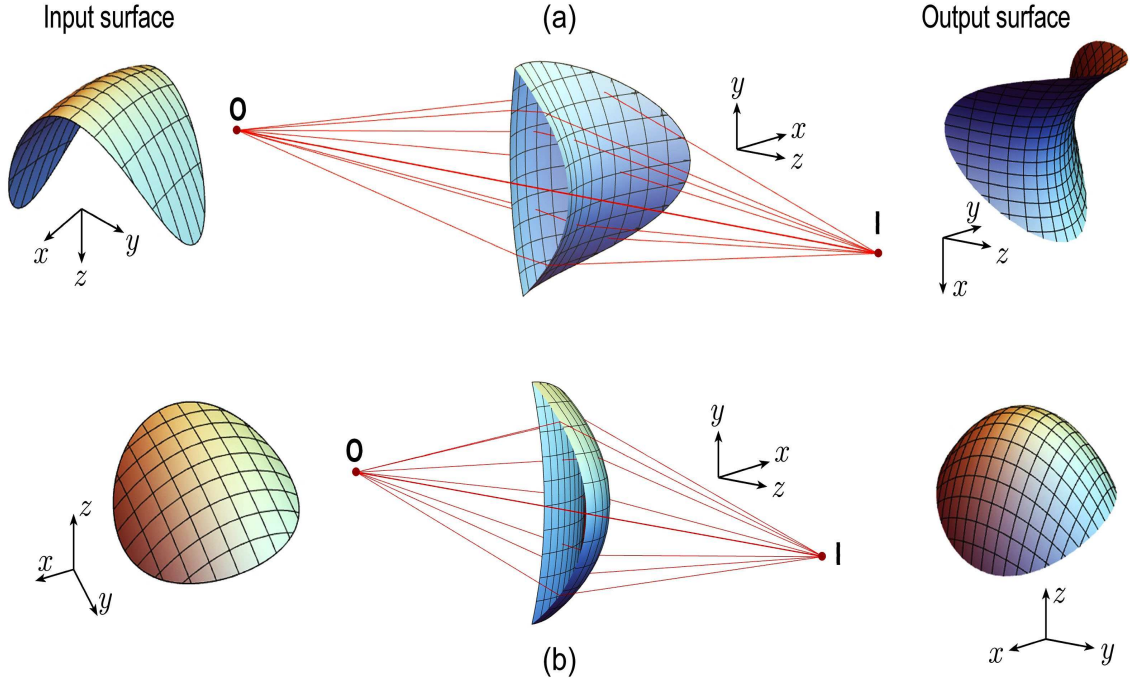}}
\caption{Singlet lenses with non-uniform (a) convex and (b) concave input surfaces with $n=1.5$, $T=1$ cm, $f_a=-5$ cm, and $f_b=6$ cm.}
\end{center}
\end{figure}
\end{center}

\section{Illustration of relevant examples}
The generality of Eqs. (\ref{sol}) allows us to show a large variety of interesting geometries of the singlet lens. In all examples the input surface is freeform and it is defined by the user. 

Let us begin by considering a non-circularly symmetric convex surface described by the elliptical paraboloid $z_a=(x_a^{2}+8y_a^{2})/200$ shown in Fig. 2(a). The output surface $z_b(x_b,y_b)$ was calculated evaluating directly Eqs. (\ref{sol}) using the design parameters included in the caption of the figure. In this case, the output function resembles a hyperbolic paraboloid. The border of the singlet is a three-dimensional curve $\textbf{r}_\textrm{border}$ that can be determined by the intersection of the input and output surfaces, i.e. $z_a(\textbf{r}_\textrm{border})=z_b(\textbf{r}_\textrm{border})$. Unfortunately, it seems that there is not a close-form analytical expression for the border, but it can be calculated numerically finding the intersection of both surfaces. Evidently the size of the lens increases as the thickness $T$ increases. Figure 2(a) also shows the trajectories for a set of rays emerging from the source (O) and converging to the image (I). For visualization purposes, we have drawn a cut of the lens as if it were hollow to appreciate clearly the surfaces and the internal trajectories of the rays. In Fig 2(b) we illustrate the case of a concave surface described by the ellipsoid $z_a=-50+(50^2-x_a^{2}-3y_a^{2})^{1/2}$.

\begin{center}
\begin{figure}
\begin{center}
\resizebox{120mm}{70mm}{\includegraphics{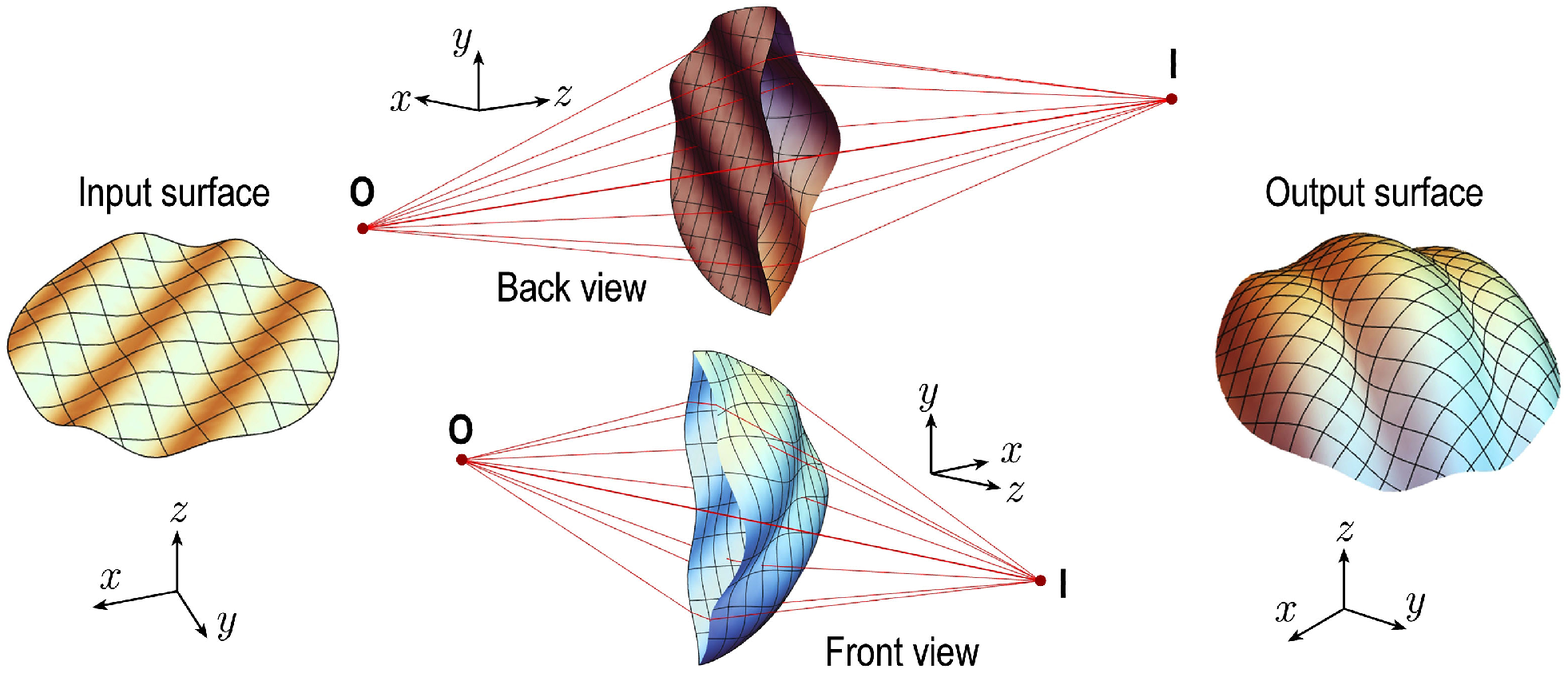}}
\caption{Singlet lens with sinusoidal input surfaces with $n=1.5$, $T=1$ cm, $f_a=-5$ cm, and $f_b=6$ cm. See Visualization 1.}
\end{center}
\end{figure}
\begin{figure}
\begin{center}
\resizebox{120mm}{70mm}{\includegraphics{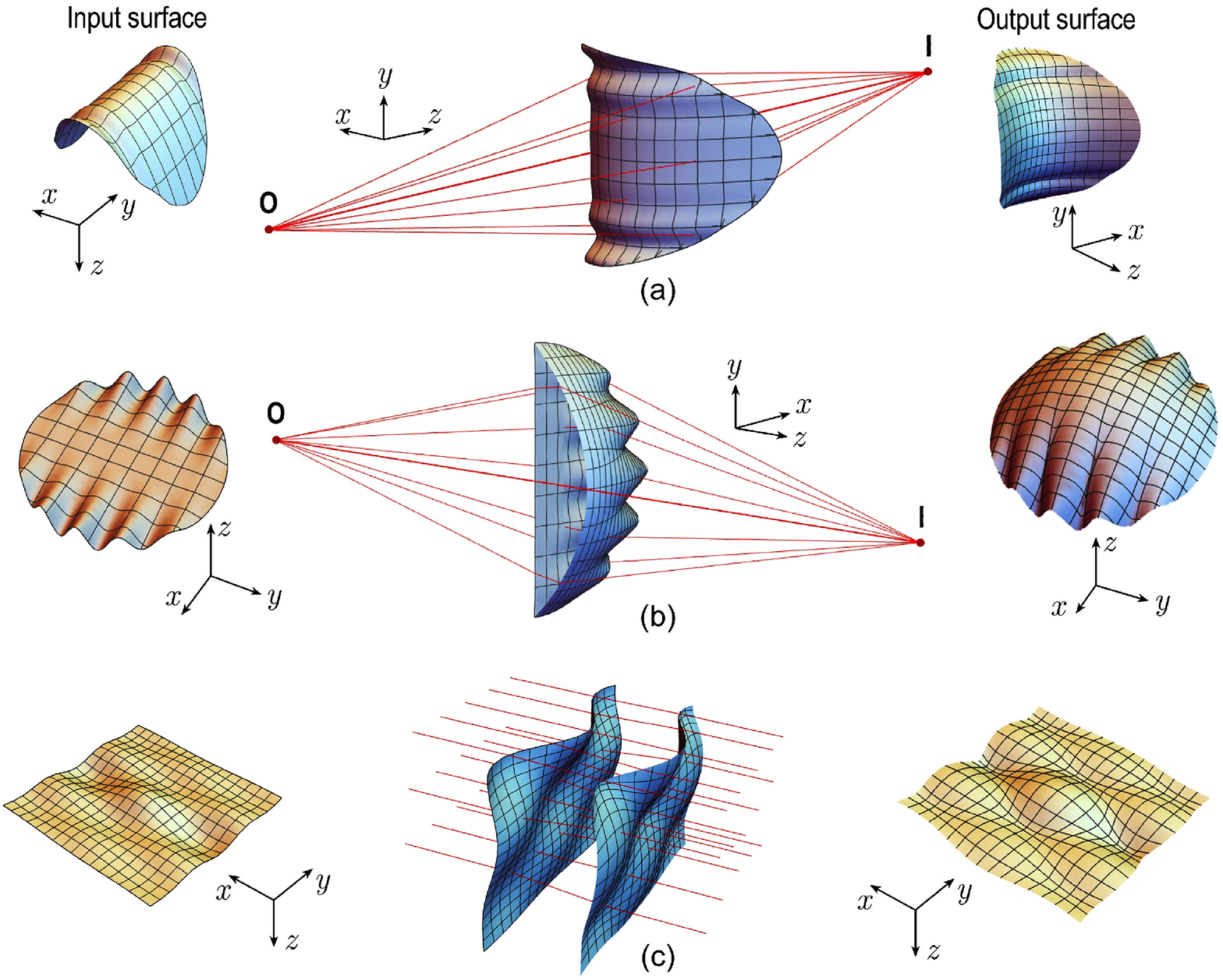}}
\caption{Singlet lenses with input surfaces mixing concave/convex with oscillating behavior. Parameters $n=1.5$, $T=1$ cm. Subplots (a) and (b) $f_a=-5$ cm, and $f_b=6$ cm. Subplot (c) $f_a \to -\infty$ and  $f_b \to \infty$. See Visualization 2 for the subplot (a).}
\end{center}
\end{figure}
\end{center}
Figure 2 shows examples of simple concave and convex input surfaces, but Eqs. (\ref{sol}) can be evaluated using more challenging input functions combining concave and convex regions. In Fig. 3 we show a singlet with an harmonic sinusoidal input surface given by $z_a=\cos(0.4x_a + 0.4y_a)$. Except for the shape of the input function, all remaining parameters are the same than in Fig. 2. The spatial frequency of the input function modulating the input surface can be increased until the limit when the rays propagating inside the glass intersect each other. We show the back and front views to facilitate the visualization of the rays traveling from (O) to (I). 

In Fig. 4 we include three additional examples of freeform lenses. In the first two subplots we illustrate the mixed behavior of convex functions with oscillating functions modulating the input surface, (a) $z_a=(x^2\cos x+8y^2)/200$, and (b) $z_a=[0.1 \cos y_a(y_a^{2} + 16x_a^{2})]/200$. From the figures, it is clear that the output rays converge to the image point despite the ripples of the input surfaces.  Finally, we remark that the positions of the object $f_a$ and the image $f_b$ can be set to any value on the optical axis including the infinity. To show this case, in Fig. 4(c) we plot the ray tracing for the input surface $z_a=-J_0(x_a)\cos(0.45 y_a)$ when the object is located at $-\infty$ and the image at $\infty$. We can see that the collimated input rays keep collimated after passing the lens but with a different distribution. Thus this device may be considered as a shaper of collimated beams.\\

\subsection{Efficiency}
To valid the efficiency of the equation (\ref{sol}), lets we compare the rays coming from the image and the rays that goes to the image, so we have the following vectors $\boldsymbol{v}_3$ and $\boldsymbol{v}_3^\dagger$, $-\boldsymbol{v}_3$ comes from the image to the second surface, and $\boldsymbol{v}_3^\dagger$ is computed using the Snell' law at the second surface, $\boldsymbol{n}_b$ is unitary the normal vector to the second surface. Therefore, $\boldsymbol{n}_b$, $\boldsymbol{v}_3$ , $\boldsymbol{v}_3^\dagger$  are written as,

\begin{equation}
\left\{\begin{array}{l}
\boldsymbol{n}_b=\pm\displaystyle{\frac{ \frac{\partial}{\partial x_a}[x_b,y_b,z_b]\times\frac{\partial}{\partial y_a}[x_b,y_b,z_b]}{\left| \frac{\partial}{\partial x_a}[x_b,y_b,z_b]\times\frac{\partial}{\partial y_a}[x_b,y_b,z_b] \right|}} ,\\\\
\boldsymbol{v}_3=\displaystyle{\frac{[r_b,z_b-T-f_b,0]}{\sqrt{r_b^2+ (z_b-T-f_b)^2}}},\\\\
\boldsymbol{v}_3^\dagger=\displaystyle{\displaystyle{n[\boldsymbol{n}_b\times (-\boldsymbol{n}_b \times \boldsymbol{v}_2)]-\boldsymbol{n}_b\sqrt{1-n^2(\boldsymbol{n}_b \times \boldsymbol{v}_2)\cdot(\boldsymbol{n}_b \times \boldsymbol{v}_2)}}}\,,
\end{array}\right.
\end{equation}
The percentage efficiency of ray is measured how close ends in the image position, therefore we defined the efficiency as
\begin{equation}
E=100\%-\left|\frac{\boldsymbol{v}_3^\dagger-\boldsymbol{v}_3}{\boldsymbol{v}_3}\right|\times 100\%
\end{equation}

We compute the efficiency for 500 rays for all the examples presented in the paper and the  average of all the examples is $99.9999999999941`\%\approx100\%$. We believe that the error is not zero because because there are computational errors such as truncation that cannot been avoided.

Please notice that for all examples the singlets are free of
spherical aberration even when the incident angles are very
large, this happens because we do not use any paraxial approximation.

We have tested a large variety of input surfaces exhibiting exotic shapes and different spatial variations. In all cases equation (\ref{sol}) gave the correct and expected behavior provided when the rays traveling inside the freeform collimator lens do not self-intersect.

\section{Conclusions}
In this paper we introduced a general formula (\ref{sol}) to design a singlet freeform lens free of spherical aberration. The method works as follows: for a given input surface $(x_a,y_a,z_a)$, the formula yields a second surface $(x_b,y_b,z_b)$ that corrects the spherical aberration generated by the first surface. We have tested many singlets with input surfaces exhibiting exotic shapes and different spatial variations for a variety of focal distances. In all cases Eqs. (\ref{sol}) gave the expected behavior provided that the rays traveling inside the freeform lens do not cross each other. In this work we have focused ourselves to eliminate the spherical aberration, but the optical systems exhibit more aberrations that we have not studied. Anyway we are convinced that this family of freeform lenses has many potential applications.

\section*{Funding}
Consejo Nacional de Ciencia y Tecnolog\'ia (Conacyt), 593740.  Tecnol\'ogico de Monterrey, 0020209I07.

\bibliographystyle{apalike}

\end{document}